%Paper: hep-th/9511088
%From: Cumrun Vafa <vafa@string.harvard.edu>
%Date: Mon, 13 Nov 95 15:32:25 -0500
%Date (revised): Tue, 14 Nov 95 10:02:15 -0500

\input harvmac.tex
\noblackbox
\Title{\vbox{\hbox{HUTP-95/A042}\hbox{\tt hep-th/9511088}}}
{Gas of D-Branes and Hagedorn Density of BPS States}
\bigskip
\centerline{Cumrun Vafa}
\bigskip\centerline{\it Lyman Laboratory of Physics}
\centerline{\it Harvard University}\centerline{\it Cambridge, MA 02138}

\vskip .3in
We test the prediction of a hagedorn density
of BPS states which carry RR charge in type II compactifications.
 We find that in certain cases they correspond to the
supersymmetric ground states for a gas of identical $0$-branes.

\Date{\it {Nov. 1995}} %replace this line by \draft  for preliminary
%%versions

%\draft

One of the key issues in understanding and testing string dualities
is the question of BPS states.  BPS states
are stable even in strong coupling regimes in the theory and
so if one finds BPS states in a given theory, it should also
appear in any equivalent dual theory.  This question becomes
even more crucial when one notices that there
is a hagedorn density of BPS states
among perturbative states in many string theories \ref\ah{
A. Dabholkar and J. Harvey, Phys. Rev. Lett. {\bf 63} (1989) 478.}.  An example
of this phenomenon
occurs for toroidal compactifications of type II strings.
In such cases there are BPS multiplets corresponding
to choosing oscillator ground states on one side
and arbitrary oscillator state on the other.

In these (and many similar) cases the corresponding proposed duals
do not exhibit the corresponding BPS states as elementary
string states and one would like
to understand how such a large degeneracy of states arises.

With the proposal of \ref\pol{J. Polchinski, preprint hep-th/9510017.}\ it has
become
clear that the relevant BPS states not manifest in the perturbative
type II spectrum should come from D-branes.  This idea has
been further developed in constructing
bound states of D-branes in \ref\wit{E. Witten, preprint hep-th/9510135.}\
leading
to a surprising connection with the study of vacua for certain supersymmetric
gauge theories.  It was shown there that the spectrum
of bound states of parallel $1$-branes and strings is consistent with the
prediction \ref\sc{J. Schwarz, preprints hep-th/9508143, hep-th/9509148,
hep-th/9510086.}\ of $SL(2,Z)$ duality of type IIB
in ten dimensions.  This development was continued in \ref\seno{A. Sen,
preprint hep-th/9510229.}\
where it was shown that even after toroidal compactification
the story for parallel D-branes is consistent with the prediction
of $U$-dualities \ref\hhht{C. Hull and P. Townsend, Nucl. Phys. {\bf B438}
(1995) 109.}.  Moreover, using the results
of \ref\BSV{M. Bershadsky, V. Sadov and C. Vafa, preprint hep-th/9510225.
}\
for intersecting D-branes it was shown \ref\sent{A. Sen,
preprint hep-th/9511026.}\ that
for the first oscillator level of BPS states the prediction
of U-duality is consistent with the bound states of intersecting
D-branes.  In this note we will extend this result to all oscillator
levels, and discuss some of its generalizations to type II-heterotic
duality in 6 dimensions.

One of the cleanest cases for testing predictions of
$U$-duality as far as BPS states is concerned is type II
 toroidal compactification
down to 6 dimensions.  This theory has 8 $U(1)$ gauge fields
coming from the NS-NS sector and 8 coming from the R-R sector.
Moreover there is an element of order two in the $U$-duality
group which exchanges these two sets of charges \ref\senv{A. Sen and C. Vafa,
preprint hep-th/9508064.}.
Let us consider the BPS states seen in the type II perturbation
theory:  The NS-NS $U(1)$ charges
of BPS state is characterized by specifying a vector
$(P_L,P_R)$ in the Narain lattice $\Gamma^{4,4}$ of the toroidal
compactification.  For each such charge the number of BPS states
is given by taking the oscillator ground states on one side (say right)
tensored with arbitrary oscillator state on the left with total
oscillator number $N_L$.  The level matching requires
\eqn\lem{N_L={1\over 2} (P_R^2-P_L^2)={1\over 2}P^2}
Moreover the mass $m$ of the BPS state is given by $m^2=P_R^2$.
Note that even if we know
the existence of a BPS state only for a given $P=(P_L,P_R)$
 the T-duality group
$SO(20,4;{\bf Z})$ implies
the existence of the same number of BPS states for any
vector in $\Gamma^{4,4}$ which is in the same orbit as $P$.

Now we will be more specific:  The $Z_2$ element of the
$U$-duality group maps these momentum states to $D$-branes
wrapped around the various cycles of $T^4$ \pol .  Moreover
the $P\cdot P$ gets mapped to the intersection number
of $p$-branes.  The case of parallel $p$-branes, where $P^2=0$
was studied in \wit (extended to the compact
case in \seno ) and was shown to agree with
the prediction of $U$-duality.
A particular case where ${1\over 2}P^2=1$
was considered in \sent\ by studying in the type IIA
two 2-branes
wrapped around 2-cycles of $T^4$, intersecting at one point in $T^4$.
Note that by $T$-duality
we could have equally well considered a 0-cycle and a
4-cycle which again intersect at one point.   This is more convenient for
us and in fact we will generalize it by considering $m$ 0-cycles
and one 4-cycle.  This gives ${1\over 2}P^2=m$.
Let us first review the result of \sent\ for $m=1$.
The expected degeneracy of BPS states in this case is $16^3$, a factor of 16
coming from right-moving ground state, another factor of 16 from
the left-moving ground state, and a factor of 16 due to the degeneracy
of the oscillators at $N_L=1$ over the left-moving ground state.
To avoid the difficulties of bound states at threshold
we first dualize type IIA to type IIB, as in \seno \sent ,
 where we end up having
one $1$-brane and one $5$-brane and consider the extra direction of the
branes wrapped an odd number of times around the extra dimension.
The resulting supersymmetric quantum field theory now lives in
different dimensions:  From the 1-brane we get the reduction
of $N=1$ $U(1)$ YM from 10 to 2 dimensions \wit , from the
5-brane, we get the reduction of $N=1$ $U(1)$ YM from 10
to 6 and from the joint configuration \BSV\ we get a matter multiplet
charged under the two $U(1)$'s living on the common 2 dimensional space.
In this case, given the triviality of the $U(1)$ dynamics,
we are justified in ignoring the dynamics of the 6 dimensional
theory and simply concentrating on the common 2-dimensional theory
where the two theories interact
which is the $d=2$ reduction
of $N=2$, $d=4$ system with  $U(1)_r\times U(1)_c/Z_2$ gauge group
with adjoint matter $\Phi^r$ and $\Phi^c$ in each $U(1)$ and
a hypermultiplet charged under $U(1)_r$. Moreover we end up
with odd electric flux in the spatial direction for the two $U(1)$'s
due to the wrapping of the D-branes around the spatial direction
\wit \sent .
 However one has
to be careful to note that since we are dealing with compact
space the bosonic component of
$\Phi^r$ and $\Phi^c$ (or its dual) parameterize $T^4$.
It was argued in
\sent\ that the gauge system for $U(1)_r$ (in the sector
with odd electric flux which cannot be screened by the
matter which carries even charge) has a
unique ground state.  Quantizing the bosonic piece of $U(1)_c$ simply
gives the space-time momenta.  Quantizing the bosonic piece of
$\Phi^r$ and $\Phi^c$ gives degeneracy one, because of the compactness
of the variables.  Quantization of the fermionic mode of
$\Phi^r$ is the same as the cohomology of $T^4$ giving
$16$ states.  The quantization of fermions in $U(1)_c$ and
$\Phi^c$ (a total of 16 fermions) gives the extra $2^8=16^2$ degeneracy
corresponding to left- and right-moving degeneracies of the ground
state.  Thus in total we get $16^3$ states as expected.
We can do more, and actually check the spacetime
quantum numbers of these states.  They will also agree
with the predictions of $U$-duality (when we recall that
the fermions in the above are spacetime spinors).

Now let us consider the case where we have $m$ 0-branes and $n$ 4-branes.
  In the type IIB setup (where we
dualize one of the directions and end up with $m$ 1-branes
and $n$ 5-branes) this would correspond to
two quantum field theories, one living in 2 dimensions with gauge group
$U(m)$ and one living in $6$ dimensions with gauge group $U(n)$
and interacting on the common $2$ dimensional space.   Unlike
the case with $n=1$ we cannot justify ignoring the non-abelian dynamics of
the $U(n)$ theory in 6-dimensions and projecting down to
2 dimensions could miss some of the states which we would have
gotten otherwise.  However, if we take the case where we have
$n=1$ but choose arbitrary $m$, then we are justified, as above
in ignoring the dynamics of the 6 dimensional theory, as it is a
free abelian theory, and simply concentrate on the common 2
dimensional theory where theory becomes interacting.
Thus we will consider the case with $m$ 0-branes and one 4-brane.
Then the system we need to consider is the $d=2$ reduction of $N=2$, d=4
with gauge symmetry
$U(m)\times U(1)_c$ with an adjoint $\Phi_1$ of $U(m)$ and
an adjoint $\Phi_2$ of $U(1)_c$, and a matter
$\Lambda$ in fundamental of $U(m)$ \foot{Amusingly enough this
is the same system that corresponds to the Calabi-Yau compactifications
with the ground ring singularity of $c=1$ string at radius $R=m$
\BSV .}.  Note that we can think of eigenvalues of $\Phi_1$ as
corresponding to the positions
of the $m$ 0-branes on $T^4$ \wit .
Note however that if the corresponding gauge theory has no mass gap,
we still have to recall that the field space is compact
because eigenvalues of $\Phi_1$ parameterize $T^4$
and we would then have in addition to quantize the
collective coordinates for the positions of the 0-branes on $T^4$.
In fact it is easy to see, using the argument of \sent\
that there is no mass gap.  The background charges will cause
the $U(1)$ of $U(m)$ to be broken, but the $SU(m)$ survives.
Since we have matter
in the fundamental of $SU(m)$ any charge at infinity
will be screened leaving no mass gap.  Another way of saying this is
as follows:
We can deform the $U(m)\times U(1)_c$
theory, by
giving expectation value to the $\Phi_1$ field so that it
breaks it to $U(1)^m \times U(1)_c$.  The fundamental
field $\Lambda$ now decompose to $m$ fields charged under
each of the $U(1)^m$'s.  Moreover
the background charge at infinity, which coupled to the
diagonal $U(1)$ now couples to all these $U(1)$'s
and thus the argument of \sent\ would show that, apart from
the choice of $\Phi_1$
we have $1^m=1$ vacuum.
However we still have to quantize the collective coordinates
which is,  after recalling the connection between $\Phi_1$ and
the points on $T^4$ and including its eight fermionic partners,
 the same as the ground states of the supersymmetric
quantum mechanical system of $m$ 0-branes on
$T^4$.  We also have to note that the $0$-branes are identical
(this also follows from the Weyl group of $U(m)$) and so the space
we are quantizing will correspond to the cohomology of $(T^4)^m/S_m$
where $S_m$ is the permutation group on $m$ objects.  This space
is singular precisely where a number (say $k$) of 0-branes coincide.  The
`resolution'
of this is dictated by the fact that the physics near these points
is governed by an enhancement to $U(k)$.  It is natural to believe
this physical `resolution' of the singularity is mirrored by
the hyperk\"ahler resolution of this space (in accord with
$N=4$ supersymmetry in $d=2$)
which agrees with the usual orbifold formula.  In fact this was
also the viewpoint adopted in
\ref\vw{C. Vafa and E. Witten, Nucl. Phys. {\bf B431} (1994) 3.
}\ in connection with similar
computations in testing the Olive-Montanen S-duality conjecture.
Actually in \vw\ the cohomology of the (resolved) $(T^4)^m/S_m$ space was
computed
where it was noted that the degeneracy of the cohomology in this
case is in one to one correspondence with the partition
function of left-movers of the superstring!  This is exactly
what we need in order to obtain the expected degeneracy of the
BPS states here (the extra $16^2$ states come from the quantization
of the fermions in $U(1)_c$ and $\Phi_2$).  Note that this
connection tells us that the $k$-th quantum of left-moving
oscillation for
the superstring corresponds to a bound state of $k$ 0-branes.
Thus the hyperk\"ahler resolution of this space in particular predicts
that even though we have no mass gap in this quantum field theory, there are
bound states of $k$ 0-branes at threshold
(which in the orbifold language come from the twisted sectors
of this theory).

 We can also
check the spacetime quantum numbers of the BPS states
and we find that they agree with these bound states.  This
again follows by the fact that the fermions of the quantum
mechanical system are spacetime spinors and the observation
(see footnote on p. 74 of
 \vw ) that  ${1\over 2}(F_L+F_R,F_L-F_R)$
 quantum numbers of the cohomology
of $(T^4)^m/S_m$ are the same as the light-cone helicities
of the left oscillator states of type II superstring compactified on
$T^4$ down to six dimensions (note that $F_L$ and $F_R$
correspond respectively to holomorphic and anti-holomorphic degrees
of the cohomology shifted by half the complex dimension).
 If we consider the case $m=1$ we get the
cohomology of $T^4$ which has sixteen elements with 6d
helicity assignments in agreement with ${1\over 2}(F_L+F_R,F_L-F_R)$.
This is sufficient to guarantee that for $(T^4)^m/S_m$ we also
get the correct helicity assignments for BPS states.

Note that if we changed the orientation of the 4-brane, the
length squared of the charge vector will become $-m$; the analysis
above continues to hold which implies we get the same degeneracy.
However this now corresponds to the BPS states, whose oscillator
states come from the right-movers.  Given the picture of
the left-oscillator states corresponding to the `gas' of 0-branes
in the presence of a 4-brane, and the right-oscillator
states corresponding to the `gas' of $0$-branes in the presence
of an anti-4-brane, it is tempting to think of strings as
`composites' of these BPS states.  This is very suggestive indeed.

A similar story seems to repeat for the type IIA compactification
on $K3$ and its duality with heterotic string on $T^4$.  In that
case the heterotic string does contain BPS states
among its elementary excitations, with the degeneracy growing
as the left-mover oscillator modes of the bosonic strings.
The corresponding analysis for the gauge degeneracy has
not been done yet, but the following observation is very suggestive:
If we consider $m$ 0-branes and one 4-brane, the requisite
degeneracy (apart from the obvious fermionic degeneracy) is
the degeneracy of open bosonic strings at level $m+1$, which
according to the result of \vw\ is the same\foot{This was the main motivation
for my conjecture
of the duality between type IIA on $K3$ and heterotic strings on $T^4$.} as
the ground states of the supersymmetric gas of $m+1$ points
on $K3$.
It remains to do the analysis of the gauge sector and verify the above
picture.

I would like to thank A. Johansen, S. Kachru, A. Sen and E. Witten for
valuable discussions.   This research is supported in part
by NSF grant PHY-92-18167.

\listrefs
\end